\documentclass{PoS}
\def\dt{\delta\tau}		

\def\hats{\hat S}		
\def\hatt{\hat T}		
\def\SST{\{S,\{S,T\}\}}
\def\TST{\{T,\{S,T\}\}}

\def\Var{\mathop{\rm Var}}      
\def\erfc{\mathop{\rm erfc}}    
\def\half{\frac12}
\def\acc{{\hbox{\tiny acc}}}
\def\PQPQP{{\hbox{\tiny PQPQPQ}}}
\newlength{\halffigwidth}
\usepackage{subfigure}
\title{Better HMC integrators for dynamical simulations}
\ShortTitle{Better HMC integrators for dynamical simulations}
\author{M. A. Clark\\
  Harvard-Smithsonian Center for Astrophysics, 
  Cambridge, MA 02138, U.S.A.\\
  E-mail: \email{mikec@seas.harvard.edu}}

\author{B\'alint Jo\'o\\
  Jefferson Lab, 12000 Jefferson Avenue, Newport News, VA 23606, U.S.A.\\
  E-mail: \email{bjoo@jlab.org}}

\author{A. D. Kennedy \\
  SUPA, School of Physics \& Astronomy, The University of Edinburgh,
  Edinburgh EH9 3JZ, Scotland, U.K.\\
  E-mail: \email{adk@ph.ed.ac.uk}}

\author{P. J. Silva\\
  Centro de F\'isica Computacional, Universidade de Coimbra, Portugal\\
  E-mail: \email{psilva@teor.fis.uc.pt}}

\abstract{We show how to improve the molecular dynamics step of Hybrid Monte
  Carlo, both by tuning the integrator using Poisson brackets measurements and
  by the use of force gradient integrators.  We present results for moderate
  lattice sizes.}

\FullConference{The XXVIII International Symposium on Lattice Field Theory\\
  June 14-19,2010\\ Villasimius, Sardinia Italy}

\begin{document}

\section{Introduction and Motivation}

Hybrid Monte Carlo (HMC) \cite{hmc} is the algorithm of choice to generate
lattice configurations including the effect of dynamical fermions.
Nevertheless, the generation of gauge field configurations at large volumes and
light quark masses is still very expensive computationally.

One principal ingredient of HMC is the molecular dynamics (MD) step, which
consists of a reversible volume-preserving approximate MD trajectory of
\(\tau/\dt\) steps (with \(\tau\) being the length of the trajectory in
suitable units, and \(\dt\) the step size) followed by a Metropolis
accept/reject test with acceptance probability \(\min(1,e^{-\delta H})\) where
\(\delta H\) is the change in the Hamiltonian \(H\) over the trajectory.

A molecular dynamics trajectory is not only an approximate integral curve of
the Hamiltonian vector field \(\hat H\) corresponding to \(H\), but is also an
exact integral curve of the Hamiltonian vector field \(\widehat{\tilde H}\) of
an exactly conserved Shadow Hamiltonian~\(\tilde H\).  The asymptotic expansion
of this Shadow Hamiltonian in the step size \(\dt\) may be computed using the
Baker--Campbell--Hausdorff (BCH) formula and expressed in terms of Poisson
brackets~\cite{lat07}.

As a simple example consider a single level Omelyan (PQPQP) integrator \cite{omelyan}
\begin{displaymath}
  U_\PQPQP(\tau) = \left(e^{\lambda\hats\dt} e^{\half\hatt\dt}
  e^{(1-2\lambda)\hats\dt} e^{\half\hatt\dt} e^{\lambda\hats\dt} \right)^{\tau/\dt}
\end{displaymath}
whose shadow Hamiltonian is 
\begin{equation}
  \tilde H_\PQPQP = H_\PQPQP + \left(\frac{6\lambda^2-6\lambda+1}{12}\SST +
  \frac{1-6\lambda}{24}\TST\right) \dt^2 + O(\dt^4).
  \label{shadow:pqpqp}
\end{equation}
Note that we have one free tunable parameter, \(\lambda\), which is often set
to some {\it ad hoc\/} value not taking Poisson brackets into
account~\cite{forcrand}.

\section{Integrator tuning}

We define the difference between the shadow (\(\tilde H\)) and actual (\(H\))
Hamiltonians as \(\Delta H=\tilde H-H\).  We have previously suggested
\cite{lat08} that \(\half\langle\delta H^2\rangle \approx\Var(\Delta H)\),
where the right hand side is the variance of the distribution of values of
\(\Delta H\) over phase space.  This formula assumes that the trajectories are
long enough that the end points are more-or-less independent of the starting
points, and accurate enough that their distribution is still close to
\(e^{-H}\).  We can therefore estimate the acceptance rate from \(\Var(\Delta
H)\)
\begin{equation}
  P_\acc = \erfc\left(\sqrt{\frac18\langle\delta H^2\rangle}\right)
  = \erfc\left(\sqrt{\frac14\Var(\Delta H)}\right)
  \label{paccvar}
\end{equation}
The advantage of using \(\Var(\Delta H)\) is that one only needs to measure the
Poisson brackets from equilibrated configurations.  We can thus express
\(P_\acc\) as a function of the integrator parameters and find their optimal
values that maximize the acceptance rate.

\begin{figure}[t]
\vspace*{-0.3cm}
  \subfigure[Acceptance rate as a function of \(\lambda\), with \(\dt=0.1\).]{
  \begin{minipage}[b]{0.45\textwidth}
    \centering
    \includegraphics[origin=c,angle=0,width=0.85\halffigwidth]%
                    {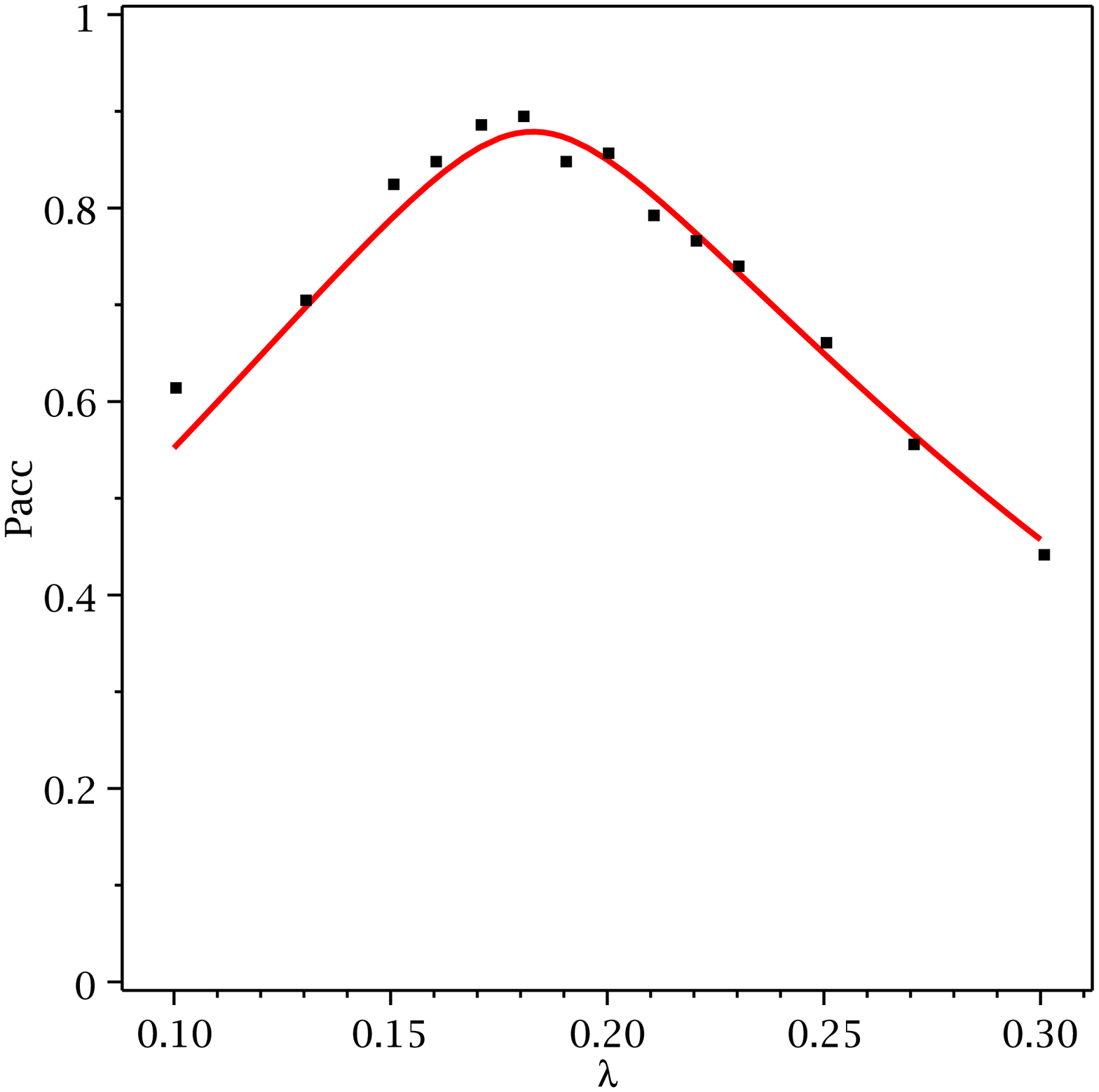}
  \end{minipage} \label{tuning:1}} \hfill
  \subfigure[Acceptance rate as a function of \(\dt\), with \(\lambda=0.18\).]{
    \begin{minipage}[b]{0.45\textwidth}
    \centering
    \includegraphics[origin=c,angle=0,width=0.85\halffigwidth]%
                    {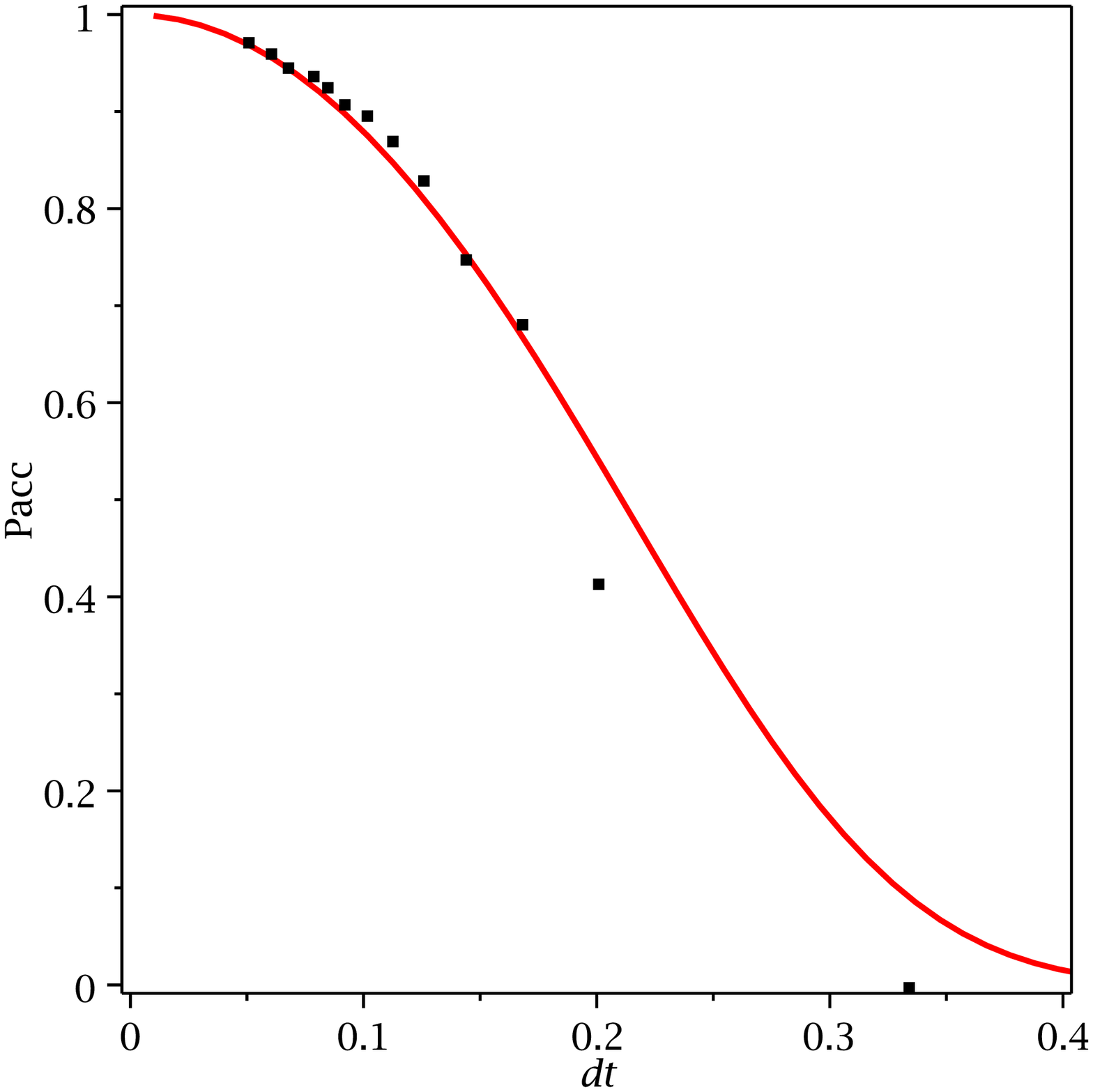}
    \end{minipage} \label{tuning:2}}
  \caption[Poisson bracket tuning]{Comparison of measured acceptance rates and
    their predictions from average Poisson brackets.}
  \label{tuning}
\end{figure}

As a simple test, we consider a HMC simulation of two flavors of Wilson
fermions at \(\kappa=0.158\) and Wilson gauge action at \(\beta=5.6\) on an
\(8^4\) lattice.  We use a single level PQPQP integrator 
and a unit trajectory length, therefore we have two parameters
to tune, namely \(\lambda\) and the step size \(\dt\).  In Figure~\ref{tuning}
we compare the acceptance rates predicted by the formula above (red curve) with
numerical data taken from simulations at various values of \(\lambda\) and
\(\dt\) (black dots).  The Poisson Bracket values used for the predictions were
measured at \(\lambda=0.18\) and \(\dt = 0.1\).  In Figure~\ref{tuning:1} we
have fixed \(\dt=0.1\) and we leave \(\lambda\) as a free parameter; whereas in
Figure~\ref{tuning:2} we take \(\lambda=0.18\) and we plot \(P_\acc\) as a
function of the step size.

The figures show good agreement between predicted and measured acceptance
rates.  We now use eq.  (\ref{paccvar}) to tune the MD integrator on a larger
volume.  Ultimately, we are interested in reducing the computational cost,
which depends on the wall-clock time spent computing the force terms as well as
the acceptance rate, and the autocorrelation time for the observables.  We
neglect the autocorrelation time in this discussion as they are not sensitive
to the choice of integrator parameters as long as the acceptance rate is
reasonable, and define our cost metric as
\begin{displaymath}
  \mbox{cost} = \frac{\mbox{trajectory CPU time}}{P_\acc}\;.
\end{displaymath}
For a nested integrator the numerator of this cost function is a function of
the number of steps at each level times the CPU time required to compute the
forces at that level.

\section{Tuning a real simulation}

As an application of our tuning technology, we are going to consider a HMC
simulation of a \(24^3\times 32\) lattice \cite{urbach}, with two flavours of
Wilson fermions with \(\kappa=0.1580\) and \(\beta=5.6\).  The authors of
\cite{urbach} include two Hasenbusch fields with twisted mass fermions as
``preconditioners'', and they use a nested PQPQP integrator scheme with
\(\lambda=1/6\) and one force term at each level, as shown in
Table~\ref{setup}, where \(0\) is the outermost level.  For each level~\(i\),
we show the corresponding number of steps~\(m_i\), the type of force and its
parameters, and typical times spent on force and force gradient computation.
All times refer to runs on \(128\) nodes on the BlueGene/L at the University
of Edinburgh.

\begin{table}[h]
  \begin{center}
    \begin{tabular}{|c|c|c|c|c|}
      \hline 
      Level \(i\) & Steps \(m_i\) & Force & F time & FG time \\
      \hline
      0 & 3 & Hasenbusch (\(\mu=0\) / \(\mu=0.057\)) & 37 s  & 43 s  \\
      1 & 1 & Hasenbusch (\(\mu=0.057\) / \(\mu=0.25\)) & 9 s &  12 s \\
      2 & 2 & Wilson (\(\mu=0.25\)) & 2.5 s & 3.3 s \\
      3 & 3 & Gauge & 0.25 s & 0.34 s \\
      \hline
    \end{tabular}
  \end{center}
  \caption[Integrator set-up]{Set-up used in the HMC simulation described
    in~\cite{urbach}, together with typical times spent on force computation.
    For convenience times for the force gradient computation used in
    \S\ref{FG-tuning} are also shown here.}
  \label{setup}
\end{table}

For simplicity, we have fixed \(\tau=1\) in our tuning exercise.  We also have
restricted our search space to \(m_0\geq3\) to avoid integrator instabilities
(which occur when the BCH expansion breaks down).

\subsection{PQPQP tuning}

We now describe how we tuned the PQPQP integrator, and what results we
obtained.  We considered two different nested schemes:
\begin{itemize}
\item[A1.] The original scheme, but with tuned values of \(\lambda\).
\item[B1.] The two Hasenbusch fields appear now at the same level (so we have
  only 3 different levels).
\end{itemize}
Table~\ref{pqpqp} shows the parameters which minimize the cost metric defined
in the last section.  For each scheme we show the number of steps at each
level, the optimal \(\lambda\) parameters, our predictions for the acceptance
rate, the estimated time spent in force computation in one trajectory, and
measurements of acceptance rates and trajectory times.  For comparison we also
show data for the original scheme.

\begin{table}[h]
  \begin{center}
    \begin{tabular}{|c|cccc|cccc|c|c|c|c|}
      \hline
      & \multicolumn4{|c|}{} & \multicolumn4{|c|}{} & 
      \multicolumn2{|c|}{Prediction} & \multicolumn2{|c|}{Measurement} \\ 
      Scheme & \multicolumn4{|c|}{\(m_i\)} &
      \multicolumn4{|c|}{\(\lambda_i\)} && \(F\) time && Time \\
      & 0 & 1 & 2 & 3 & 0 & 1 & 2 & 3 & \(P_\acc\) & / traj. &
      \(P_\acc\) & / traj. \\
      \hline
      Original & 3 & 1 & 2 & 3 & \(1/6\) & \(1/6\) & \(1/6\) & \(1/6\) &
      \(0.85\) & 655 s & \(0.89\) & 709 s \\
      A1 & 3 & 1 & 1 & 2 & \(0.185\) & \(0.188\) & \(0.184\) & \(0.183\) &
      \(0.80\) & 541 s & \(0.75\) & 578 s \\ 
      B1 & 3 & 3 & 1 & $-$ & \(0.177\) & \(0.183\) & \(0.176\) & --- & \(0.82\)
      & 454 s & \(0.83\) & 554 s \\
      \hline
    \end{tabular}
  \end{center}
  \caption[PQPQP tuning]{Tuning of the PQPQP integrator scheme.}
  \label{pqpqp}
\end{table}

We see that both the tuning of the integrator parameters and changes to the
scheme provide further improvements over an already well-tuned scheme.
Indeed, with the B1 scheme we get a 1.3\(\times\) speedup.

\subsection{Force gradient integrator tuning} \label{FG-tuning}

We have again considered the two integrator schemes above, but used a PQPQP
force gradient integrator \cite{lat09} at all levels.  As in this case there
are no tuneable parameters, we could only vary the number of steps at each
level.  In Table~\ref{fg2} we show the best parameters we found as well as the
measured values of acceptance rates and trajectory times.

\begin{table}[h]
  \begin{center}
    \begin{tabular}{|c|cccc|c|c|c|c|}
      \hline
      & \multicolumn4{|c|}{} & \multicolumn{2}{|c|}{Prediction} &
      \multicolumn{2}{|c|}{Measurement} \\ 
      Scheme & \multicolumn4{|c|}{\(m_i\)} && \(F+FG\) && Time \\
      & 0 & 1 & 2 & 3 & \(P_\acc\) & time / traj. & \(P_\acc\) & / traj. \\
      \hline
      Original & 3 & 1 & 2 & 3 &  \(0.85\) & 655 s & \(0.89\) & 709 s \\
      A2 & 3 & 1 & 1 & 1 & \(0.96\) & 778 s & \(0.91\) & 814 s \\
      B2 & 3 & 1 & 1 & $-$ & \(0.91\) & 565 s & \(0.78\) & 626 s \\
      \hline
    \end{tabular}
  \end{center}
  \caption[FG tuning]{Tuning of the force gradient integrator scheme.}
  \label{fg2}
\end{table} 

In this case, the acceptance rates are not sufficiently high to amortize the
higher wall-clock time cost coming from the computation of the force gradient
term.  Thus overall, we could see no measurable improvement in the cost metric
as compared to the regular PQPQP case.  However, further improvement could be
possible by either using other force gradient integrators \cite{omelyan}, or
tuning the Hasenbusch masses.  We are currently working on these issues.

\section{Conclusions} 

We have presented a novel way of tuning an integrator, together with a
practical example using a moderate lattice size.  This tuning procedure can be
used for all lattice gauge and fermionic actions.  We are working towards a
general implementation of the calculation of Poisson brackets and force
gradient terms in Chroma~\cite{chroma}.  In the near future we will consider
the tuning of HMC simulations on larger lattices and smaller quark masses, and
we will also consider other widely used lattice actions.

\acknowledgments

P.~J.~Silva acknowledges support from FCT via grant SFRH/BPD/40998/2007, and
project PTDC/FIS/100968/2008.  B\'alint~Jo\'o acknowledges funding through 
US D.O.E Grants DE-FC02-06ER41440, DE-FC02-06ER41449 (SciDAC) and 
DE-AC05-060R23177 under which Jefferson Science Associates LLC manages
and operates the Jefferson Lab.  M.~A.~Clark
acknowledges support from NSF via award No. PHY-0835713.  The U.S. Government
retains a non-exclusive, paid-up, irrevocable, world-wide license to publish
or reproduce this manuscript for U.S. Government purposes.

The numerical results have been obtained using Chroma library~\cite{chroma}.

\end{document}